\documentclass[reprint,superscriptaddress,amsmath,amssymb,aps,prab]{revtex4-1}

\usepackage{graphicx}
\usepackage{natbib}
\usepackage{hyperref}
\usepackage{cleveref}
\usepackage[caption=false,labelformat=empty]{subfig}
\usepackage{mathrsfs}
\usepackage{gensymb}
\usepackage{upgreek}

\hypersetup{colorlinks=true,citecolor=blue,urlcolor=blue}

\newcommand{\rmd}{\mathrm{d}}
\newcommand{\micron}{{\upmu\mathrm{m}}}

\newcommand{\Energy}{\mathcal{E}}

\newcommand{\abs}[1]{\left| #1 \right|}
\newcommand{\Ecrit}{E_\text{cr}}
\newcommand{\avg}[1]{{\left\langle {#1} \right\rangle}}

\begin{document}

\title{Model-independent inference of laser intensity}

\author{T. G. Blackburn}
\email{tom.blackburn@physics.gu.se}
\affiliation{Department of Physics, University of Gothenburg, SE-41296 Gothenburg, Sweden}
\author{E. Gerstmayr}
\author{S. P. D. Mangles}
\affiliation{The John Adams Institute for Accelerator Science, Imperial College London, London SW7 2AZ, United Kingdom}
\author{M. Marklund}
\affiliation{Department of Physics, University of Gothenburg, SE-41296 Gothenburg, Sweden}

\date{\today}

\begin{abstract}
An ultrarelativistic electron beam passing through an intense laser pulse emits radiation around its direction of propagation into a characteristic angular profile.
Here we show that measurement of the variances of this profile in the planes parallel and perpendicular to the laser polarization, and the mean initial and final energies of the electron beam, allows the intensity of the laser pulse to be inferred in way that is independent of the model of the electron dynamics.
The method presented applies whether radiation reaction is important or not, and whether it is classical or quantum in nature, with accuracy of a few per cent across three orders of magnitude in intensity.
It is tolerant of electron beams with broad energy spread and finite divergence.
In laser-electron beam collision experiments, where spatiotemporal fluctuations cause alignment of the beams to vary from shot to shot, this permits inference of the laser intensity at the collision point, thereby facilitating comparisons between theoretical calculations and experimental data.
\end{abstract}

\maketitle

\section{Introduction}

Electromagnetic fields of extraordinary strength are produced at the focus of modern
high-power lasers~\cite{danson.hplse.2019}, inducing nonlinear classical, even quantum, dynamics of particles and
plasmas~\cite{mourou.rmp.2006,marklund.rmp.2006,dipiazza.rmp.2012}.
The amplitude of these fields, expressed covariantly through the normalized vector potential $a_0$,
plays an essential role in determining which regime is explored.
However, it remains difficult to diagnose \emph{in situ} the intensity reached in experiments.
This is particularly acute for experiments at or beyond the current intensity frontier,
which will explore the transition to the nonlinear quantum regime~\cite{neitz.prl.2013,blackburn.prl.2014}.
As the dynamics are not fully understood, it is important to know what $a_0$ is reached
for comparison between competing theoretical descriptions and experimental data.
Furthermore, the method by which $a_0$ is determined should not be sensitive to the
underlying physics,
particularly if the latter is the subject of the experiment.

The method presented here is based on the collision of an ultrarelativistic electron beam
with the laser pulse; this geometry has been already been exploited as
a high-energy photon source~\cite{chen.prl.2013,sarri.prl.2014,yan.np.2017}
and a probe of radiation reaction~\cite{cole.prx.2018,poder.prx.2018}.
Measurement of the angular profile of the resulting radiation
has been proposed as a means of determining the peak intensity of a laser
pulse~\cite{harshemesh.ol.2012,harvey.prab.2018}
(the former demonstrated experimentally in \cite{yan.np.2017}),
as has measurement of the electron scattering angles~\cite{mackenroth.njp.2019}.
However, the results presented in \cite{harshemesh.ol.2012,harvey.prab.2018,mackenroth.njp.2019}
depend critically on the model assumed for the electron dynamics.
The appropriate choice of model depends on the intensity of the laser to be probed:
for example, at very high intensity, radiation reaction and quantum effects
are expected to become important, if not dominant~\cite{dipiazza.rmp.2012}.
A method that does not require such an assumption to be made
would be a useful complement to methods that are model-dependent,
providing stronger evidence that a particular regime has been reached.

Here we show that the laser intensity can be inferred in a model-independent
way, using the angular profile of the radiation emitted in the collision
of the laser with an electron beam, in combination with the mean initial
and final energies of the beam.
We derive analytical predictions for the size of the radiation profile
and the energy change of the electron beam,
treating the laser as a pulsed plane electromagnetic wave,
that can be combined so as to eliminate an explicit dependence
on classical radiation reaction.
We show that this model-independence applies to a high degree of accuracy under
quantum models of radiation reaction as well.

Examining our method in a more realistic scenario, where the tight focussing
of the laser and finite size of the electron beam are taken into account,
we find that it yields a model-independent estimate of the laser intensity, averaged over the electron-beam size.
Furthermore, it is robust against finite energy spread and angular divergence, two important characteristics of non-ideal electron beams; the latter, in particular, controls the overlap between laser and electron beam.
This permits measurement of the peak intensity, if the electron beam is well-characterized, stable, and radially smaller than the laser focal spot size; with spatiotemporal fluctuations taken into account, our method provides a powerful constraint on the intensity at the point of interaction, for each individual collision.
This is complementary to methods aimed at determining the peak intensity itself,
by measurement of the ionization level of heavy atoms~\cite{ciappina.pra.2019},
Thomson scattering of low-energy electrons present in the focal volume~\cite{he.oe.2019},
or by detailed characterization of the laser structure, gathered over
hundreds of shots~\cite{jeandet.jpp.2019}.
In conjunction with these,
our method provides a means of determining the shot-to-shot overlap between
laser pulse and electron beam.

\section{Analytical results}

Consider an electron (charge $-e$ and mass $m$)
with Lorentz factor $\gamma \gg a_0$ oscillating in a
linearly polarized, plane electromagnetic wave that has normalized amplitude $a_0$
and frequency $\omega_0$. Over a single cycle of the wave, the angle between the
electron momentum and the laser axis is $\theta(\phi) = a_0 \sin\phi / \gamma$,
and the electron's quantum parameter $\chi(\phi) = 2 \gamma a_0 \omega_0 \abs{\cos\phi} / m$.
Here $\phi$ is the phase and the angle
$\theta$ lies in the plane defined by the laser's electric field and wavevector;
we refer to angles in this plane as being \emph{parallel} ($\parallel$) to the
laser polarization.
We work throughout in natural units where $\hbar = c = 1$.
The distribution of energy radiated per unit angle
$\rmd \Energy_\text{rad}/\rmd \theta$ may be calculated by integrating
the Larmor power, which is proportional to $\chi^2(\phi)$, over the cycle
and assuming that the radiation is strongly beamed along the electron's
instantaneous momentum, i.e. at phase $\phi$ the emission angle is
$\theta(\phi)$. We then normalize the result by the total radiated
energy to obtain
$\rmd \Energy_\text{rad}/\rmd \theta = 2 \gamma \sqrt{1 - (\gamma \theta/a_0)^2} / (\pi a_0)$
for $\abs{\theta} < a_0/\gamma$. The variance of the distribution is
$\sigma_\parallel^2 = \int \theta^2 \,\rmd\Energy_\text{rad} = a_0^2 / (4\gamma^2)$.

To incorporate a pulse envelope $f(\phi)$ and the effect of radiation reaction
into this result, we use the fact that the variances introduced each cycle may be added linearly.
The contribution of each cycle of the pulse to the total variance is
$a_0^2 f^2(\phi) /[4\gamma^2(\phi)]$, weighted by $\gamma^2(\phi) f^2(\phi)$.
The weighting comes from the Larmor power, which is proportional to the square
of the instantaneous Lorentz factor $\gamma(\phi)$ and the local electric field,
which is proportional to $f(\phi)$.
$\gamma(\phi)$ is obtained by solution of the Landau-Lifshitz equation~\cite{landau.lifshitz},
which accounts for the self-consistent radiative energy loss.

The total variance, in the direction parallel to the laser polarization,
can be expressed compactly in terms of the final Lorentz factor
$\gamma_f$ and integrals over the pulse envelope:
	\begin{equation}
	\sigma_\parallel^2 =
		\frac{a_0^2}{4 \gamma_i \gamma_f}
		\frac{\int\! f^4(\phi) \,\rmd\phi}{\int\! f^2(\phi) \,\rmd\phi}
		+ \sigma_\perp^2.
	\label{eq:ParallelVariance}
	\end{equation}
The second term in \cref{eq:ParallelVariance} accounts for the
contributions of the intrinsic divergence of the radiation and any initial
divergence of the electron beam $\delta$, which we assume to be
cylindrically symmetric.
These are the only contributions in the direction \emph{perpendicular}
to the laser polarization and wavevector~\cite{blackburn.pra.2020}:
	\begin{equation}
	\sigma_\perp^2 =
		\frac{5}{8\gamma_i \gamma_f} + \delta^2.
	\label{eq:PerpVariance}
	\end{equation}
In both \cref{eq:ParallelVariance,eq:PerpVariance}, we have~\cite{dipiazza.lmp.2008}
	\begin{align}
	\gamma_f &= \frac{\gamma_i}{1 + R \gamma_i},
	&
	R &=
		\frac{2\alpha a_0^2 \omega_0}{3 m}
		\int_{-\infty}^\infty\! f^2(\phi) \,\rmd\phi,
	\label{eq:FinalGamma}
	\end{align}
where $\alpha = e^2/(4\pi)$ is the fine-structure constant.
In the absence of radiation reaction, $\alpha = 0$ and $\gamma_i = \gamma_f$.
If the intensity profile $f^2(\phi)$ is a Gaussian with full-width-at-half-maximum
(FWHM) duration $\tau$, we have $\int\! f^2(\phi) \,\rmd\phi = \omega_0 \tau \sqrt{\pi/(4\ln{2})}$ and
$\int\! f^4(\phi) \,\rmd\phi = [\int\! f^2(\phi) \,\rmd\phi]/\sqrt{2}$.
Notice that the radiation profile is elongated along the polarization direction;
strictly, this result is valid for $a_0 \gtrsim 1$, as we have in this work, otherwise
the profile would be dipolar in shape~\cite{seipt.pra.2013}.

\begin{figure}
	\includegraphics[width=\linewidth]{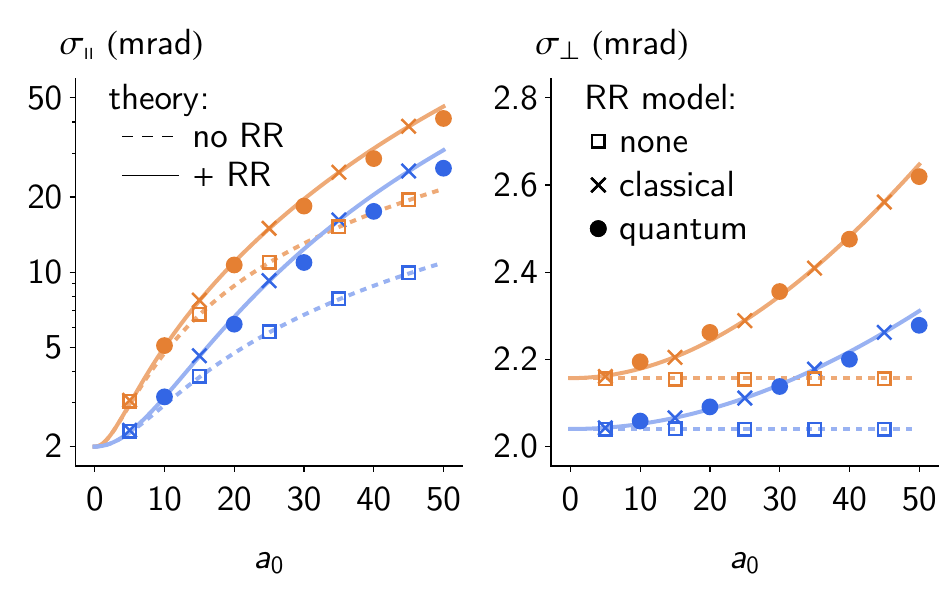}
	\caption{
		Parallel and perpendicular standard deviations of the angular profile of the
		radiation emitted by an electron beam with mean initial energy
		500~MeV (orange) and 1000~MeV (blue), as predicted by
		\cref{eq:ParallelVariance,eq:PerpVariance} with (solid) and without RR (dashed),
		and from simulations with the specified radiation reaction model (points).
		}
	\label{fig:Sigma}
	\end{figure}

A comparison of \cref{eq:ParallelVariance,eq:PerpVariance} to the results of
simulations is shown in \cref{fig:Sigma}. In these, the plane-wave laser pulse has
a Gaussian temporal envelope with FWHM duration $\tau = 40$~fs and a wavelength of
$0.8~\micron$. The energies of the beam electrons are normally distributed,
with a mean of 500 or 1000~MeV and standard deviation 50~MeV
in both cases; the initial divergence is $\delta = 2$~mrad.
Three models of the dynamics are considered: no radiation reaction (RR), i.e. the
Lorentz force only; classical RR in the Landau-Lifshitz prescription;
and a fully stochastic, quantum model using probability rates calculated
in the locally constant field approximation~\cite{ritus.jslr.1985}.
These models are described in detail in \cref{app:Models}.

Radiation is generated by Monte Carlo sampling of the classical synchrotron
spectrum in the former two cases and the quantum synchrotron spectrum in
the latter, as is appropriate.
This method is applicable at high intensity and at high harmonic order
in both the classical~\cite{reville.apj.2010,wallin.pop.2015} and
quantum regimes~\cite{ridgers.jcp.2014,gonoskov.pre.2015}, where
the photon formation length becomes much smaller than the wavelength of the
driving laser~\cite{ritus.jslr.1985}.
Such photons dominate the power spectrum, which is the object of analysis in this work,
even for relatively low $a_0$~\cite{blackburn.pop.2018}.
We use a synchrotron spectrum that is differential in both energy and
angle in our particle-tracking code, thereby resolving the intrinsic
divergence of the radiation around the electron's instantaneous velocity
vector~\cite{baier.book,blackburn.pra.2020}.
Expressions for the emission spectrum are given in \cref{app:Models} for each model.

We find excellent agreement between the analytical predictions and the simulation
results for classical and no RR. It is clear that that radiation reaction leads
to increased broadening of the angular profile in the plane of polarization.
Furthermore, the classical and quantum models give generally similar results,
even though the broadening is smaller for the latter because the radiated power
is reduced.
In the plane perpendicular to the laser polarization, the variance is
dominated by the initial divergence of the electron beam and radiation reaction
effects are weaker.

	\begin{figure}
	\includegraphics[width=\linewidth]{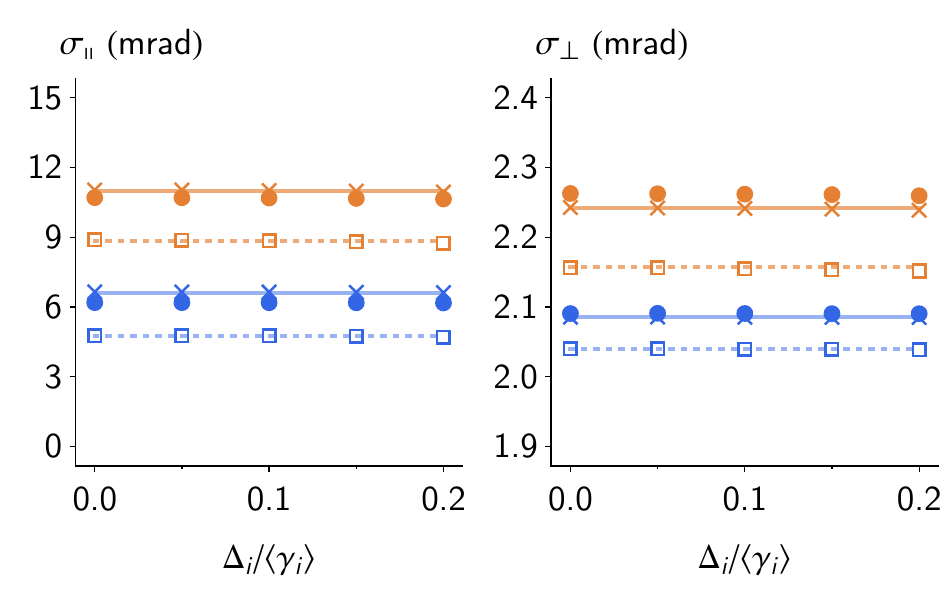}
	\caption{
		Parallel and perpendicular standard deviations as a function of initial energy
		spread $m \Delta_i$, for an electron beam with mean initial energy
		$m \avg{\gamma_i} = 500$~MeV (orange) and 1000~MeV (blue)
		colliding with a laser pulse with $a_0 = 20$:
		simulation results
		with quantum (filled circles), classical (crosses) and no (squares) radiation reaction (as in \cref{fig:Sigma}),
		and theoretical predictions
		\cref{eq:ParallelVariance,eq:PerpVariance} with (solid)
		and without (dashed) RR, assuming that the beam is monoenergetic
		at the mean energy.
		}
	\label{fig:EnergySpread}
	\end{figure}

The effect of a finite energy spread on the variances is surprisingly small.
Consider a beam of electrons, in which the initial Lorentz factors $\gamma_i$
are distributed as $\frac{\rmd N_e}{\rmd\gamma_i}$.
The variance of the radiation angular profile, in the direction parallel to
the laser polarization, for the beam as a whole, is obtained
by integrating \cref{eq:ParallelVariance}, weighted by
$(\gamma_i - \gamma_f) \frac{\rmd N_e}{\rmd\gamma_i}$,
over all $\gamma_i \gg 1$. The weighting reflects the fact
that electrons with higher $\gamma_i$ radiate more energy
and therefore contribute more to $\sigma^2_\parallel$.
We obtain
	\begin{equation}
	\sigma_\parallel^2 =
		\frac{a_0^2}{4\sqrt{2}}
		\left[ \avg{\gamma_i} \avg{\gamma_f} + \mathrm{cov}(\gamma_i,\gamma_f) \right]^{-1}
		+ \sigma_\perp^2,
	\label{eq:ApproxParallelVariance}
	\end{equation}
where $\avg{\gamma_{i,f}}$ are the mean initial and final Lorentz factors.
The covariance term $\mathrm{cov}(\gamma_i,\gamma_f) \leq \Delta_i \Delta_f \leq \Delta_i^2$, where the $\Delta_{i,f}$ are the 
standard deviations of $\gamma_{i,f}$; the equality holds when radiation
reaction is neglected. Even if $\Delta_i$ is as large as $\avg{\gamma_i}/3$, i.e. the beam has
close to 80\% energy spread (full width at half max),
the change in $\sigma_\parallel$ is at most 5\%
and we can safely neglect any effect of the initial energy spread.
As an example, we show in \cref{fig:EnergySpread} that $\sigma_\parallel$ and
$\sigma_\perp$ are unchanged when the energy spread is increased from $0$ to
$0.2$ of the mean initial energy.

\section{Intensity inference}
\subsection{Plane waves}
\label{sec:1D}

We now show that the angular profile and \cref{eq:ApproxParallelVariance}
can be used to obtain the laser intensity, i.e. the collision $a_0$, in a
model-independent way. The key points are that the analytical result is
given in terms of the mean initial and final energies and that the
covariance term is negligible.
This being the case, $a_0$ is fixed by the mean initial and final electron energies,
which can be measured; any explicit dependence on
radiation reaction effects is absorbed into the latter quantity.
Rearranging \cref{eq:ApproxParallelVariance}, we obtain
	\begin{equation}
	a_0^2 =
		4\sqrt{2} \avg{\gamma_i} \avg{\gamma_f}
		( \sigma_\parallel^2 - \sigma_\perp^2 ).
	\label{eq:InferredIntensity}
	\end{equation}
There is no explicit dependence on classical radiation reaction
because we have taken the \emph{product} of the final energy and
the angular size:
while $\gamma_f \propto (1 + R \gamma_i)^{-1}$, as shown by \cref{eq:FinalGamma},
$( \sigma_\parallel^2 - \sigma_\perp^2 ) \propto (1 + R \gamma_i)$.
Heuristically, because the electron oscillation
in a electromagnetic wave is proportional to $a_0 / \gamma$,
a decrease in the energy causes a compensating increase in
the size of the electron oscillation and consequently the angular
spread of the radiation.
Furthermore, taking the difference between the parallel and perpendicular
variances removes the effect of the electron beam divergence $\delta$,
as this term appears in both \cref{eq:ParallelVariance,eq:PerpVariance}.
However, it should be noted that the divergence has an importance not evident in the 1D scenario considered in this section, which is that it controls the expansion of the electron beam and therefore its degree of overlap with the laser focal spot.
This is discussed in \cref{sec:3D}.

\begin{figure}
	\includegraphics[width=\linewidth]{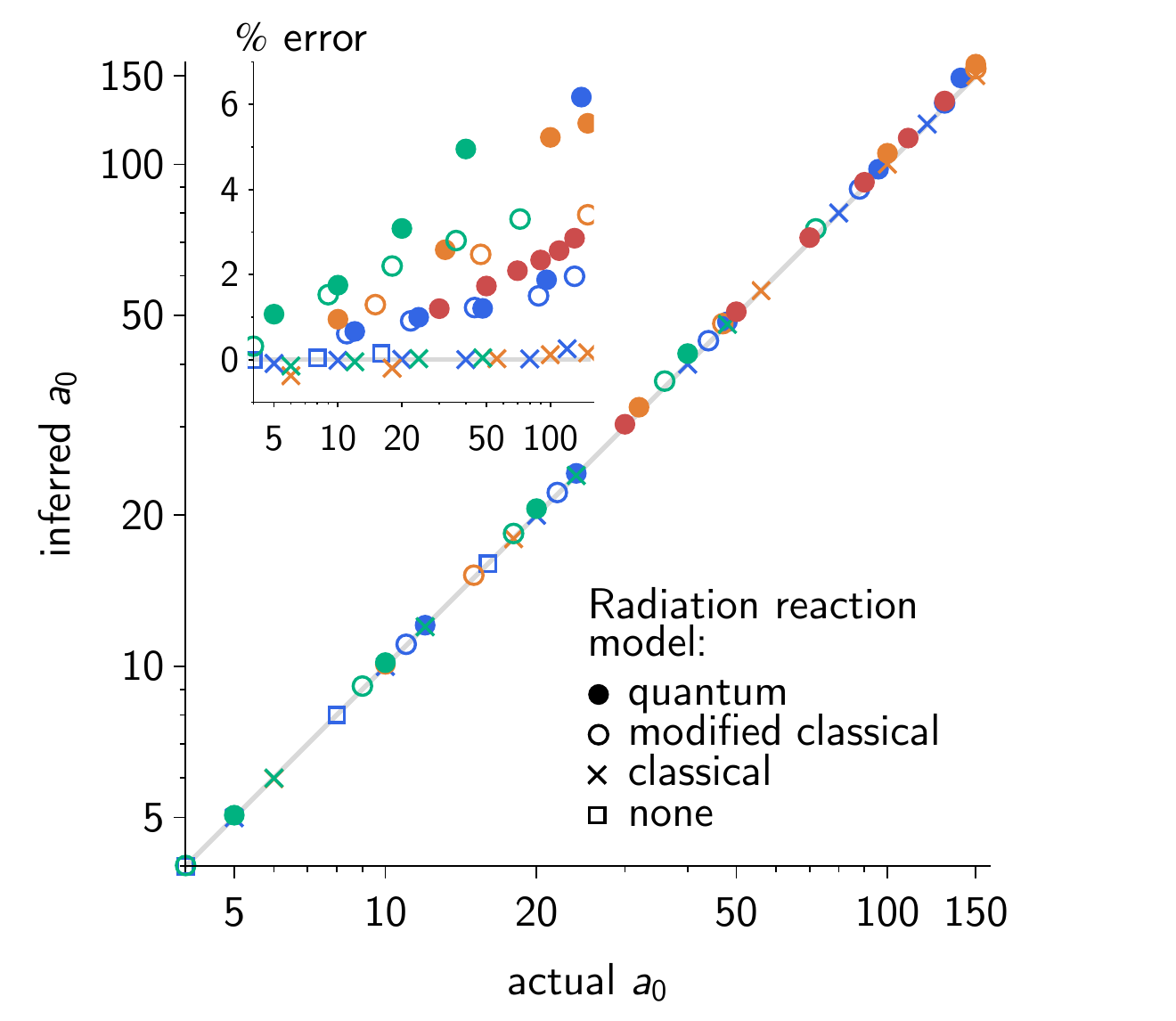}
	\caption{
		The $a_0$ inferred from simulation results, using \cref{eq:InferredIntensity},
		and (inset) the percentage error in the same,
		for an beam of electrons with normally distributed initial
		energies (mean $\mu$ and standard deviation $\Delta$)
		and initial divergence $\delta$
		colliding with a plane-wave laser pulse with normalized amplitude $a_0$,
		wavelength $0.8~\micron$ and FWHM duration $\tau$:
		(blue) $\mu = 250$~MeV, $\Delta = 5$~MeV, $\delta = 1$~mrad, $\tau = 150$~fs;
		(orange) $\mu = 500$~MeV, $\Delta = 50$~MeV, $\delta = 3$~mrad, $\tau = 20$~fs;
		(green) $\mu = 1000$~MeV, $\Delta = 50$~MeV, $\delta = 2$~mrad, $\tau = 30$~fs;
		and (red) $\mu = 200$~MeV, $\Delta = 10$~MeV, $\delta = 2$~mrad, $\tau = 30$~fs.
		}
	\label{fig:InferIntensity}
	\end{figure}

Although \cref{eq:InferredIntensity} is exactly true under classical
radiation reaction (and its absence), we now demonstrate that it
works very well under quantum models of radiation reaction as well.
We determine what $a_0$ must have been for a set of simulations,
substituting into \cref{eq:InferredIntensity}
the $\sigma_\parallel$, $\sigma_\perp$, and
the mean initial and final electron energies obtained from those
simulations~\footnote{The mean $\mu$ and standard deviation $\sigma$
of an angular probability
density function $p(\theta)$ are defined by the relations
$\tan\mu = \avg{\sin\theta}/\avg{\cos\theta}$ and
$\sigma^2 = -2 \ln [\avg{\sin\theta}^2 + \avg{\cos\theta}^2]$, where
$\avg{f(\theta)} = \int_0^{2\pi}\! f(\theta) p(\theta) \,\rmd\theta$.}.
We vary the energy spectra of the electron beams, their initial divergence,
the laser intensity and duration, and the model of RR used to calculate the dynamics.
The laser pulse is modelled as a plane EM wave.
In addition to the three models introduced earlier (and described in \cref{app:Models}), we also consider
a modified classical model, in which radiation losses are continuous,
but the strength of the Landau-Lifshitz force is reduced by the Gaunt
factor $g(\chi) \leq 1$~\cite{erber.rmp.1966},
and the photon energies are sampled from the quantum synchrotron
spectrum (see Supplemental Material).
This has been used to describe recent experimental results~\cite{poder.prx.2018}.
It ensures that the power spectrum is quantum-corrected, but neglects
stochastic effects.
(See \cref{app:Models} for details.)
The inferred $a_0$ is plotted against the true $a_0$ in \cref{fig:InferIntensity}.

We find that using \cref{eq:InferredIntensity} to infer the laser $a_0$ is
accurate to within a few per cent across the range of parameters explored,
whether radiation reaction is classical or quantum in nature, or absent.
If no radiation reaction, or classical RR, is used in the simulations,
the agreement is near perfect; when a quantum-corrected model is used
instead, the error grows with increasing electron beam energy and $a_0$.
This suggests that, while $\chi$-dependent corrections can be made to
\cref{eq:ParallelVariance,eq:PerpVariance}, most of their effect
is encapsulated in the dependence on the final energy $\gamma_f$.
For example, in a collision between an electron beam with
mean energy 500~MeV (standard deviation 5~MeV,
orange points in \cref{fig:InferIntensity})
and a laser pulse with $a_0 = 100$, going from the classical to the quantum model
increases the mean final energy by a factor of $1.58$, but decreases
the parallel variance $\sigma_\parallel^2$ by a factor of $1.43$. Therefore
the inferred $a_0 = 105$ is close to the actual $a_0 = 100$, which is
the classical result.

For $a_0 > 50$, it is advisable to use electron beams of lower energy
to minimize the importance of quantum corrections to \cref{eq:InferredIntensity}:
reducing the mean initial energy to 200~MeV (red points in \cref{fig:InferIntensity})
from 1000~MeV (green points), reduces the error in the inferred $a_0$
by more than a factor of two.
The difficulty associated with doing so is that a detector with larger
acceptance angle is required to capture the whole radiation profile,
which has characteristic size $\propto a_0/(\gamma_i \gamma_f)^{1/2}$.
Our simulation results indicate that capturing all photons with
$\theta_\text{max} \lesssim 2 a_0 / \gamma_f$ is necessary for accurate determination
of $\sigma_\parallel^2$.
This angle is almost independent of $\gamma_i$ at high $a_0$ due to
radiative losses, where it grows as $a_0^3$.
These radiative losses also mean that we do not necessarily have $\gamma \gg a_0$
throughout the laser pulse, as assumed in our earlier derivation.
As such, we expect the method presented here to be limited by the reflection threshold
$\gamma_f \simeq a_0$,
above which the re-acceleration of the decelerated electrons by the
laser pulse becomes significant~\cite{li.prl.2015,vranic.sr.2018}.
In fact, quantum effects intercede before this is reached.
Nevertheless, the region $5 \lesssim a_0 \lesssim 150$ shown in
\cref{fig:InferIntensity} is relevant for a wide variety of laser-electron
scattering experiments at existing, and planned, high-intensity laser facilities.

In analyzing the simulation results, we have used the fact that
the mean initial energy  for each individual shot $\avg{\gamma_i}$ is known exactly.
While the \emph{final} energy can be measured straightforwardly
on a shot by shot basis, it is unlikely that this can be done for
the initial energy, unless the advanced method proposed by
\cite{baird.njp.2019} can be employed.
Therefore the $\avg{\gamma_i}$ appearing in \cref{eq:InferredIntensity}
must be obtained from measurements of the electron beam in the
absence of the high-intensity laser.
However, our results in \cref{fig:IntensityScan} indicate that
the accuracy of the inferred $a_0$ is a few per cent,
which means, if the stability of the mean initial electron
energy is better than 10\%, this is not the dominant
source of error.

\subsection{Focussed fields}
\label{sec:3D}

We now consider the application of our results in a more realistic
configuration, where we take into account the spatiotemporal structure of a
tightly focussed laser pulse and the finite size of the electron beam.
It is clear that, ideally, the electron beam should be much smaller than the
laser focal spot, in the same way that any probe must be smaller than the
system to be probed.
If this is not the case, the inferred $a_0$ will be smaller than the true
$a_0$, as the radiation profile will have been averaged over the
spatial profile of the electron beam.
This is still a useful quantity, as it represents an $a_0$ that is
characteristic of the collision as a whole.
Indeed, we show that if a transverse offset is introduced between
electron beam and laser pulse, the reduction in the inferred $a_0$
allows the imperfect overlap to be identified.

	\begin{figure}
	\includegraphics[width=\linewidth]{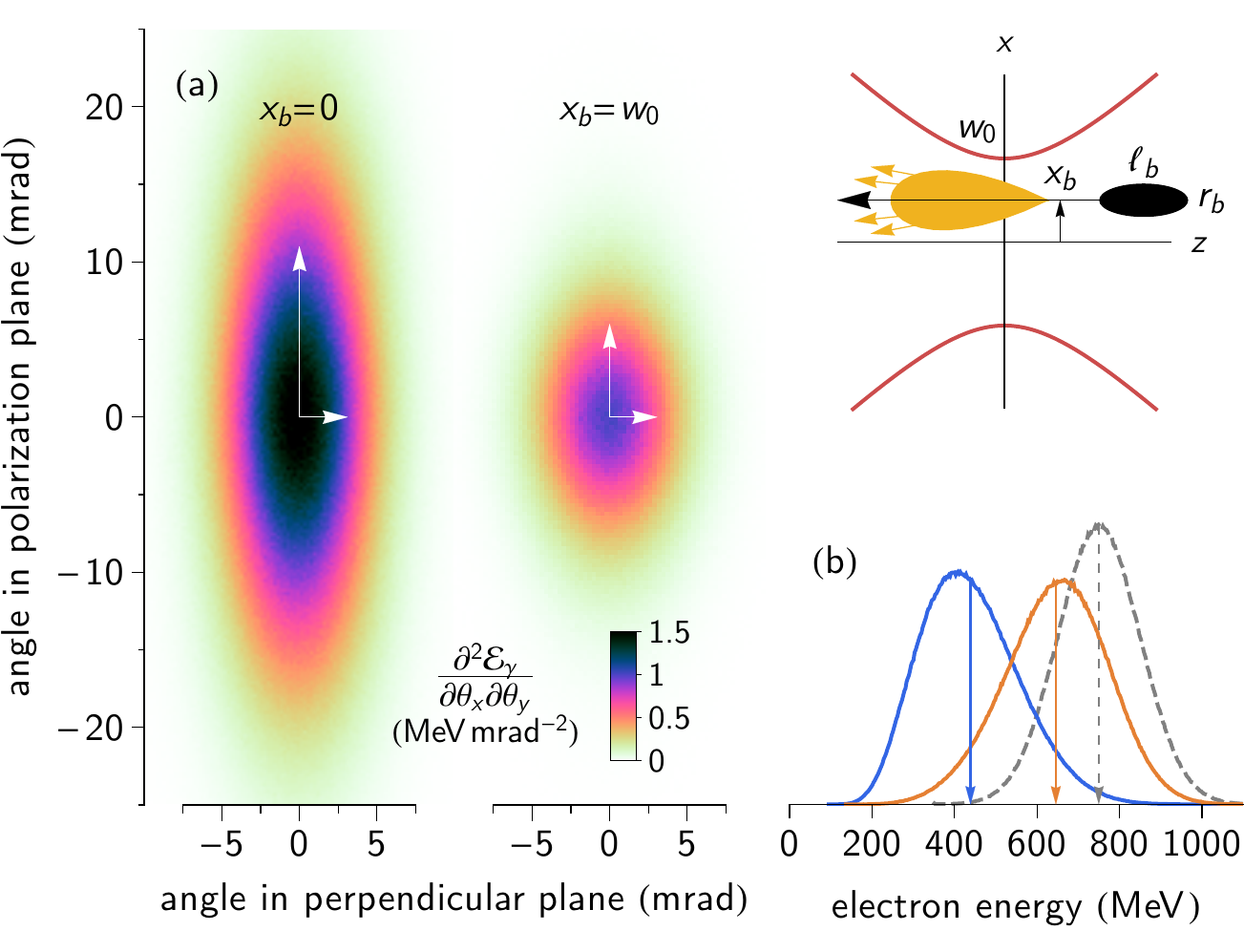}
	\caption{
		(a) Energy radiated per unit solid angle (per electron)
		by an electron beam in a collision with a laser pulse that has peak
		$a_0 = 30$.
		White arrows indicate the standard deviations of the distributions.
		(b) Energy spectra of the electrons before the collision (grey, dashed)
		and after, when the beam offset from the laser axis is $x_b = 0$ (blue)
		and $x_b = w_0$ (orange). Arrows indicate the mean energy.
		}
	\label{fig:TestSpectra}
	\end{figure}

Consider the collision of an electron beam with a laser pulse that has
wavelength $0.8~\micron$ and FWHM duration 30~fs, which is focussed to
a spot size of $w_0 = 2~\micron$, where $w_0$ is the radius at which
the intensity falls to $1/e^2 \simeq 0.14$ of its peak value.
The electron beam has a cylindrically symmetric, Gaussian charge
distribution of radius $r_b = 0.5~\micron$ and length $\ell_b = 5.0~\micron$;
it has mean energy 750~MeV (standard deviation 100~MeV),
rms divergence 3~mrad, and is offset from the laser axis by a perpendicular distance $x_b$.
The angular distributions of
the emitted photons and electron energy spectra for this configuration are shown in \cref{fig:TestSpectra},
for $a_0 = 30$ and assuming quantum radiation reaction.
The inferred $a_0$ are $a_0^\text{inf} = 28.2$ and $16.7$ for $x_b = 0$ and
$w_0$ respectively. The former is consistent with the true value of $a_0$;
the reduction in the latter case is evidence
that electron beam has not interacted with the most intense part of
the laser pulse. Notice that the average energy loss of the electrons
and the angular profile of the radiation are both reduced in size.

\begin{figure}
	\includegraphics[width=\linewidth]{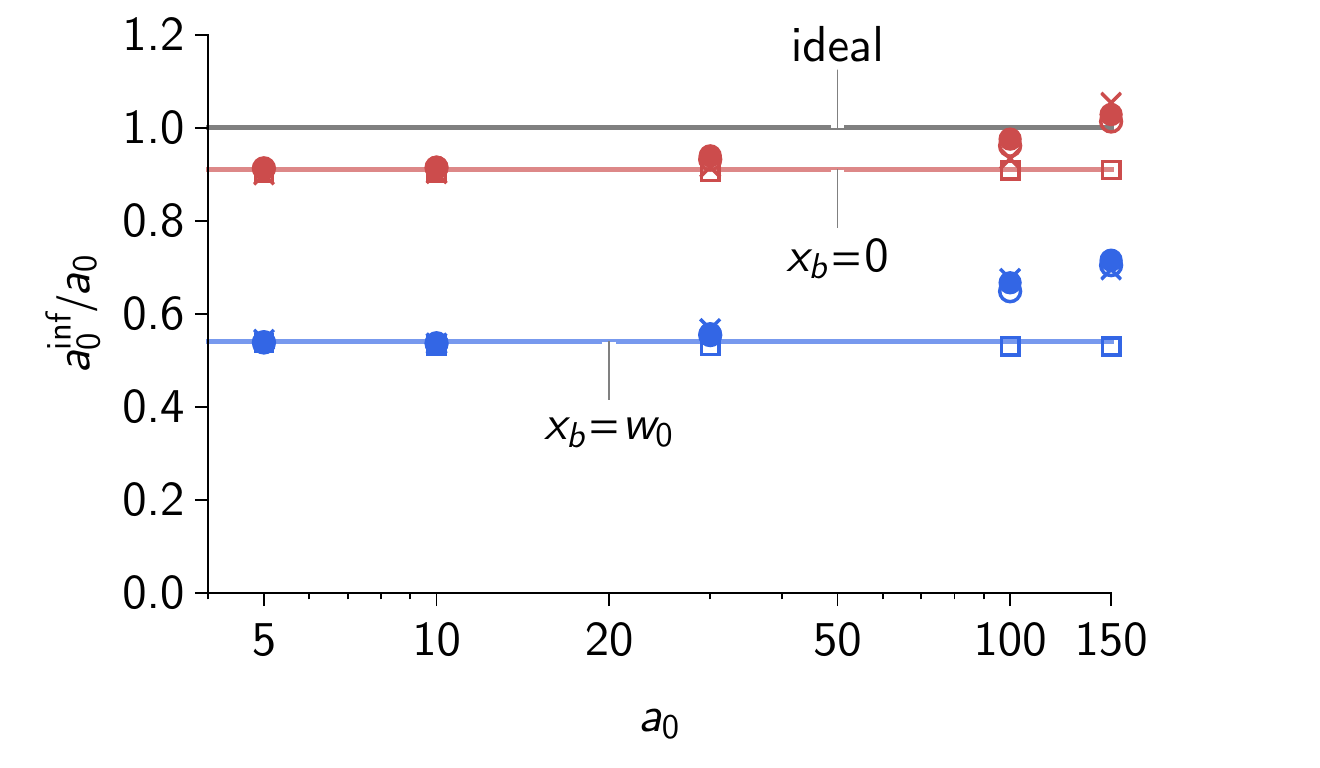}
	\caption{
		The inferred $a_0$ as a fraction of the true $a_0$, for the
		collision geometry shown in \cref{fig:TestSpectra},
		for different values of $a_0$:
		from simulations with quantum (filled circles),
		modified classical (open circles), classical (crosses)
		and no RR (squares);
		and as predicted by \cref{eq:FiniteSizeEffects} for
		$x_b = 0$ (red) and $x_b = w_0$ (blue).
		}
	\label{fig:IntensityScan}
	\end{figure}

We can estimate the reduction due to finite-size effects as follows.
An individual electron of the beam, with transverse displacement $x,y$,
encounters a peak normalized laser amplitude $a(x,y) = a_0 \exp[-(x^2+y^2)/w_0^2]$.
As the radiation profile is an integrated signal, the $a_0$ inferred from it is
$a_0^\text{inf} = \sqrt{\avg{a^4}/\avg{a^2}}$ to lowest order in $\alpha$,
where the average is taken over the distribution of $x,y$.
Then
    \begin{equation}
    a_0^\text{inf} = a_0 \sqrt{P/Q} \exp[-\xi^2/(PQ)],
    \label{eq:FiniteSizeEffects}
    \end{equation}
where $P = 1 + 4\rho^2$, $Q = 1 + 8\rho^2$, $\rho = r_b/w_0$ and $\xi = x_b/w_0$.
First we confirm that \cref{eq:InferredIntensity} provides a
model-independent prediction of the laser $a_0$ by repeating the
simulations shown in \cref{fig:TestSpectra} for different models of
radiation reaction: \cref{fig:IntensityScan} shows that the inferred $a_0$
is consistent across all four models tested, at a peak $a_0 < 150$.
We also find that the reduction in the inferred $a_0$ due to the finite
size of the electron beam and the transverse offset is in agreement with
\cref{eq:FiniteSizeEffects}.

For the highest $a_0$ shown in \cref{fig:IntensityScan}, the
three radiation-reaction models, while consistent with each other,
separate from the `no RR' result.
This is due to ponderomotive scattering, which
is the radial expulsion of electrons from a focussed field
by intensity gradients, and therefore a source of angular
deflection absent in a plane wave.
Such deflection is amplified by radiation losses, which reduce $\gamma$ and
so the rigidity of the electron beam.
Consequently, the radiation profile is broader for a focussed field than
for a plane wave with the same peak $a_0$ and \cref{eq:InferredIntensity}
overestimates the intensity.
The error grows to 10\% at $a_0 \simeq 150$, for a focal spot size
$r_0 = 2.0~\micron$. Nevertheless, taking this as the upper bound
on the validity of the method we have presented, we conclude that it does infer
$a_0$ in a model-independent way across approximately three orders
of magnitude in laser intensity, as shown in \cref{fig:InferIntensity}.

	\begin{figure}
	\includegraphics[width=\linewidth]{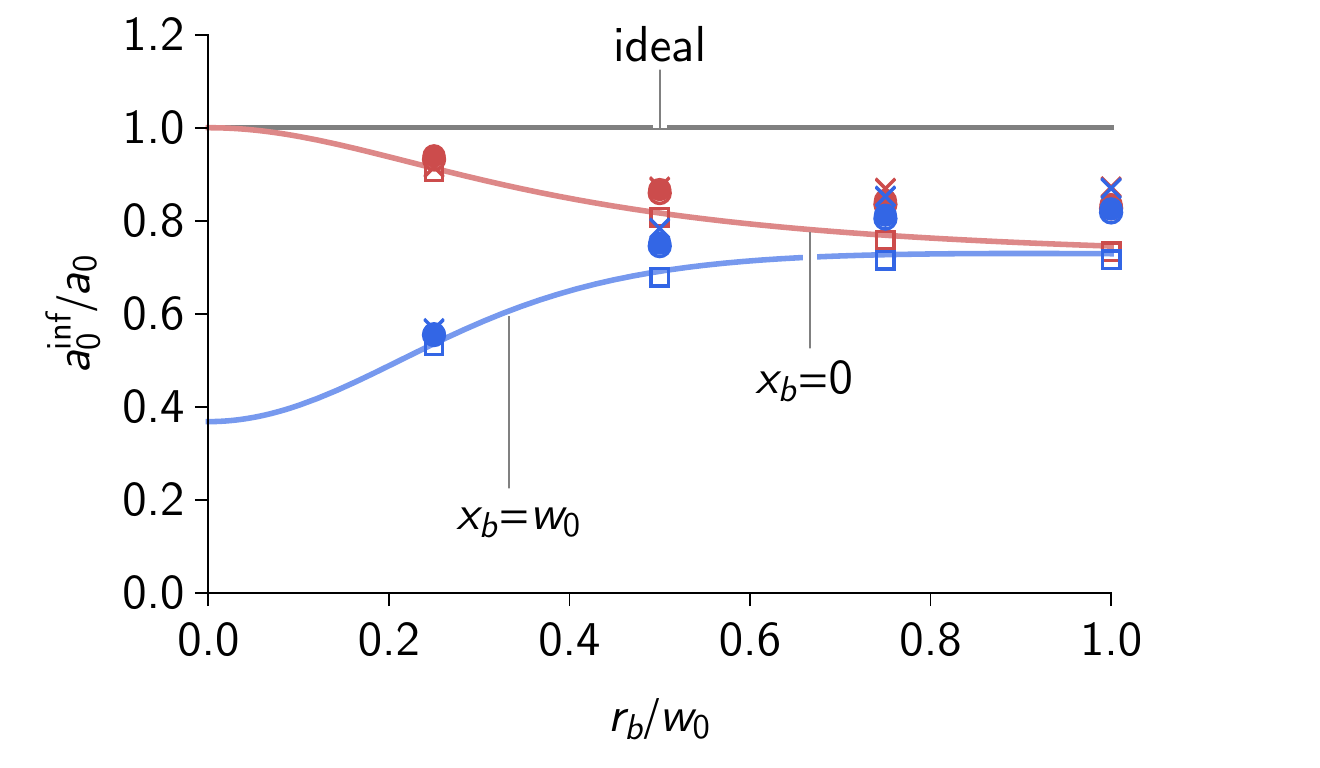}
	\caption{
		The inferred $a_0$ as a fraction of the true $a_0$, for the
		collision geometry shown in \cref{fig:TestSpectra},
		for different beam radii $r_b$:
		from simulations with quantum (filled circles),
		modified classical (open circles), classical (crosses)
		and no RR (squares);
		and as predicted by \cref{eq:FiniteSizeEffects} for
		$x_b = 0$ (red) and $x_b = w_0$ (blue).
		}
	\label{fig:RadiusScan}
	\end{figure}

In the simulations underpinning \cref{fig:TestSpectra,fig:IntensityScan}
we set the beam radius $r_b = 0.5~\micron$, which corresponds to a
full-width-at-half-maximum diameter of $1.2~\micron$.
While beam sizes of this magnitude have been measured in
laser-wakefield accelerators~\cite{schnell.prl.2012},
a large distance between the end of the acceleration stage
and the collision point will cause the beam size to be larger.
In \cref{fig:RadiusScan}, we show how the inferred $a_0$ depends
on the beam radius at fixed $a_0 = 30$, in the cases that there
is no transverse offset between the beams ($x_b = 0$, in red) and
an offset of $2.0~\micron$ ($x_b = w_0$, in blue).
We increase the beam radius $r_b$, while keeping the divergence
$\delta$ fixed, as if the distance between the beam generation
and collision points had been increased.
Our results confirm that increasing the beam size
until it is comparable to the size of the laser focal spot has a
relatively weak effect on the inferred $a_0$.
Moreover, it has the additional effect of making the procedure
more tolerant of a transverse offset between the beams.
The reduction in $a_0^\text{inf}$ from $a_0$ is good agreement
with \cref{eq:FiniteSizeEffects}.

Given a separate estimate, or measurement, of the peak laser
intensity, the transverse offset $x_b$ could be inferred from
the reduction of $a_0^\text{inf}$ from $a_0$.
Furthermore, as the overlap between laser pulse and electron beam
varies from shot to shot due to imperfect pointing stability,
accumulating the \emph{distribution} of $a_0^\text{inf}$ over a
large series of collisions could indicate systematic effects
such as the finite size of the electron beam.

\section{Conclusions}

We have shown that by probing an intense laser pulse
with a relativistic electron beam, measuring the angular size
of the emitted radiation and the initial and final beam energies,
the normalized amplitude $a_0$ at the collision point may be
inferred in a model-independent way. By `model-independent',
we mean that across three orders of magnitude in laser intensity,
$5 \lesssim a_0 \lesssim 150$,
different models for the electron dynamics yield a consistent
value for the inferred $a_0$ that is accurate to within a few per cent.
As the best choice of model depends on the intensity,
relaxing the requirement that one must be assumed
\textit{a priori} means that our method provides strong evidence
that a particular intensity range has been reached.
This is particularly useful for experiments intended to
distinguish between radiation reaction models, as this
becomes feasible in a reduced number of successful collisions
if $a_0$ can be measured independently~\cite{arran.ppcf.2019}.
The quantities necessary to use \cref{eq:InferredIntensity},
our main result, can be measured on a shot by shot basis
without additional theoretical modelling.
This allows the variation in the collision $a_0$ due to shot-to-shot
fluctuations to be identified, including the effect of a
systematic error in alignment.
We emphasize that our method is complementary to model-dependent
analysis of the interaction, using, for example, the largest
angle of the radiation angular profile~\cite{harshemesh.ol.2012},
or coincidence measurements of the radiation and electron energy
spectra~\cite{cole.prx.2018}.

The datasets necessary to reproduce the figures and analysis
are available in Ref.~\cite{dataset}.

\begin{acknowledgments}
T.G.B. thanks Arkady Gonoskov for helpful discussions.
We acknowledge funding from
the Knut and Alice Wallenberg Foundation (T.G.B. and M.M.),
the Swedish Research Council (grant 2016-03329, M.M.),
and the Engineering and Physical Sciences Research Council
(grant EP/M018555/1, S.P.D.M.).
This work was supported by the European Research Council (ERC)
under the European Union's Horizon 2020 research and innovation
programme grant agreement 682399 (S.P.D.M.).
Simulations were performed on resources provided by the
Swedish National Infrastructure for Computing (SNIC)
at the High Performance Computing Centre North (HPC2N).
\end{acknowledgments}

\appendix
\section{Models of radiation reaction and emission}
\label{app:Models}

We use \texttt{circe}, a particle-tracking code that simulates photon and
positron production in prescribed external electromagnetic waves.
Collective effects and back-reaction are neglected in this framework.
There are four possible models of the electron dynamics:
no radiation reaction (i.e. Lorentz force only);
classical radiation reaction using the Landau-Lifshitz equation;
a modified classical model incorporating a quantum correction to the radiated power;
and a quantum, stochastic model.
Photon emission is handled in the same way for all four models,
except that in the former two the photon momentum is sampled from the
classical synchrotron spectrum, and in the latter two from the
quantum synchrotron spectrum.
Here we discuss the four models in detail.

\subsection{No radiation reaction}

In the \emph{no RR} model, the electron trajectory follows from the
Lorentz force equation:
	\begin{equation}
	\dot{p}^\mu = -\frac{e F^{\mu\nu} p_\nu}{m},
	\label{eq:LorentzForce}
	\end{equation}
for four-momentum $p$ and field tensor $F$.
(Dots denote differentiation with respect to proper time.)
The photon emission rate $\dot{N}_\gamma$, differential in
photon energy $\omega$, polar angle $\theta$ and azimuthal angle $\varphi$,
is controlled by the electron Lorentz factor $\gamma = (1-\beta^2)^{-1/2}$,
velocity $\beta$, and quantum parameter $\chi = \abs{F_{\mu\nu} p^\nu}/ (m \Ecrit)$,
where $\Ecrit = m^2/e$ is the critical field of QED~\cite{schwinger.pr.1951}.
In the classical limit $\chi \ll 1$, it is given by~\cite{baier.book,esarey.pre.1993}:
	\begin{equation}
	\frac{\partial^3 \dot{N}_\gamma}{\partial u \partial z \partial\varphi} =
		\frac{\alpha m}{3\sqrt{3}\pi^2 \chi}
		u (2z^{2/3} - 1)
		K_{1/3}\!\left( \frac{2 u z}{3 \chi} \right).
	\label{eq:ClassicalEmissionRate}
	\end{equation}
Here $u = \omega / (\gamma m)$, $z = [2\gamma^2(1 - \beta\cos\theta)]^{3/2}$
and $K$ is a modified Bessel function of the second kind.
The domain of \cref{eq:ClassicalEmissionRate} is $0 \leq u < \infty$,
$1 \leq z < \infty$ and $0 \leq \varphi < 2\pi$.

The trajectory is obtained numerically using the following methods.
If the external field is a plane wave, the particle push takes the following
form~\cite{blackburn.pop.2018}:
the spatial components of the momentum $p^\mu$ perpendicular to the laser
wavevector $\kappa$ are determined by
$\omega_0 \partial_\phi \vec{p}_\perp = -e \vec{E}_\perp(\phi)$,
where $\vec{E}_\perp$ is the electric field at phase $\phi$
and the angular frequency $\omega_0 = \kappa^0$.
The other two components follow from the conditions $p^- = \text{const}$
and $p^2 = m^2$, and the position from $\omega_0 \partial_\phi x^\mu = p^\mu / p^-$.
Here $p^- = \kappa.p/\omega_0$ is the \emph{lightfront} momentum.
If the field is a focussed Gaussian beam, and therefore a function of all
three spatial coordinates, we use the particle push introduced by Vay~\cite{vay.pop.2018}
and the analytical expressions given in~\cite{salamin.apb.2007}.

To model photon emission, each electron is initialized with an
optical depth $T = -\log(1-U)$ for pseudorandom
$0 \leq U < 1$, which evolves as $\dot{T} = -\dot{N}_\gamma$,
where $\dot{N}_\gamma$ is the instantaneous rate of emission
(per unit proper time) obtained by integrating \cref{eq:ClassicalEmissionRate},
until the point where $T$ falls below zero.
Then $T$ is reset and the photon energy $\omega$, polar scattering angle $\theta$
and azimuthal angle $\varphi$ are pseudorandomly sampled
from the differential spectrum \cref{eq:ClassicalEmissionRate}.

\subsection{Classical radiation reaction}

The equation of motion in the \emph{classical} model is
the Landau-Lifshitz equation~\cite{landau.lifshitz}, which
adds to \cref{eq:LorentzForce} a continuous radiation-reaction
force $f_\text{LL}$ that accounts for the loss of energy:
	\begin{align}
	\dot{p}^\mu &= -\frac{e F^{\mu\nu} p_\nu}{m} + f^\mu_\text{LL},
	\label{eq:LandauLifshitz}
	\\
	f^\mu_\text{LL} &= \frac{2\alpha m}{3}
		\!\left[
			-\frac{F^{\mu\nu}_{,\sigma} p_\nu p^\sigma}{m^2 \Ecrit}
			+ \frac{F^{\mu\nu} F_{\nu\sigma} p^\sigma}{\Ecrit^2}
			- \chi^2 p^\mu
		\right].
	\label{eq:LandauLifshitzForce}
	\end{align}
The photon emission rate associated with the resulting
trajectory is exactly as given in \cref{eq:ClassicalEmissionRate}.

Numerically, if the external field is a plane wave, the lightfront momentum, which satisfies
$\omega_0 \partial_\phi p^- = -(2\alpha/3) |\vec{E}_\perp(\phi)/E_\text{cr}|^2 {p^-}^2$,
is advanced using Heun's method. The remaining components follow from
the mass-shell condition and
$\omega_0 \partial_\phi (\vec{p}_\perp / p^-) = -e \vec{E}_\perp(\phi) / p^-$,
where we have neglected an additional derivative term that is smaller
by a factor of $\alpha \omega_0/m \ll 1$.
If the field is a focussed Gaussian beam, we employ the Vay push as before,
modified to include RR using the method given in \cite{tamburini.njp.2010}:
the momentum after half a timestep is used to calculate the magnitude of
the RR force $\vec{f}_\text{LL}$, and the associated momentum change
$\vec{f}_\text{LL} \Delta t$ is added to the momentum change induced by
the Lorentz force.
We use only the last term in \cref{eq:LandauLifshitzForce} to calculate
$\vec{f}_\text{LL}$, as this is by far the dominant contribution.

Radiation emission is modelled in exactly the same way as in the
\emph{no RR} model.
The difference is that the energy radiated, according to \cref{eq:ClassicalEmissionRate},
matches the energy lost by the electron.

\subsection{Quantum radiation reaction}

In the quantum picture, radiation reaction is the recoil arising
from the emission of multiple, incoherent photons.
At high intensity, it is possible to model emission as occurring
instantaneously and discretely at stochastically determined points.
Thus the equation of motion
\emph{between} emission events is given by the Lorentz force, \cref{eq:LorentzForce}.
At an emission event, the electron recoil is determined
self-consistently by the momentum of the photon it has emitted,
i.e. $\vec{p} \to \vec{p} - \vec{k}$.

Therefore we obtain the trajectory numerically using the same methods
as in the \emph{no RR} case.
Photon emission is modelled by sampling the quantum synchrotron
spectrum~\cite{baier.book}:
	\begin{multline}
	\frac{\partial^3 \dot{N}_\gamma}{\partial u \partial z \partial\varphi} =
		\frac{\alpha m}{3\sqrt{3}\pi^2 \chi}
		\frac{u}{(1+u)^3}
	\\
		\times \left[
			z^{2/3} (2 + 2u + u^2) - (1 + u)
		\right]
		K_{1/3}\!\left( \frac{2 u z}{3 \chi} \right)
	\label{eq:EmissionRate}
	\end{multline}
where $u = \omega / (\gamma m - \omega)$, $z = [2\gamma^2(1 - \beta\cos\theta)]^{3/2}$
and $K$ is a modified Bessel function of the second kind.
(The domain of \cref{eq:EmissionRate} is $0 \leq u < \infty$,
$1 \leq z < \infty$ and $0 \leq \varphi < 2\pi$.)
Note that, unlike \cref{eq:ClassicalEmissionRate}, there is no emission
of photons with more energy than the electron.
As a consequence, the radiated power is reduced, with respect
to its classical value, by a factor $g(\chi)$~\cite{erber.rmp.1966}.

As in the \emph{no RR} and \emph{classical} models,
each electron is initialized with an
optical depth $T = -\log(1-U)$ for pseudorandom
$0 \leq U < 1$, which evolves as $\dot{T} = -\dot{N}_\gamma$,
where $\dot{N}_\gamma$ is the instantaneous probability rate of emission
(per unit proper time) obtained by integrating \cref{eq:EmissionRate},
until the point where $T$ falls below zero.
Then $T$ is reset and the photon energy $\omega$, polar scattering angle $\theta$
and azimuthal angle $\varphi$ are pseudorandomly sampled
from the differential spectrum \cref{eq:EmissionRate}.
The electron momentum after the emission $p'$ is fixed by
the conservation of three-momentum, $\vec{p}' = \vec{p} - \vec{k}$,
which induces an error that is small for ultrarelativistic particles~\cite{duclous.ppcf.2011}.

\subsection{Modified classical radiation reaction}

A well-known deficiency of the classical model is that it
predicts the emission of photons with more energy than
the electron. This can be corrected phenomenologically
by using the quantum emission spectrum \cref{eq:EmissionRate},
rather than the classical equivalent \cref{eq:ClassicalEmissionRate},
and weakening the Landau-Lifshitz force \cref{eq:LandauLifshitzForce}
by the Gaunt factor $g(\chi)$~\cite{erber.rmp.1966}.
Concretely, the equation of motion becomes
	\begin{equation}
	\dot{p}^\mu = -\frac{e F^{\mu\nu} p_\nu}{m} + g(\chi) f^\mu_\text{LL},
	\label{eq:ModifiedLandauLifshitz}
	\end{equation}
where $f_\text{LL}$ is as defined in \cref{eq:LandauLifshitzForce}.
In this way, radiation reaction is still manifest as a continuous
force, the energy radiated matches the energy lost, and the
first quantum correction is accounted for~\cite{ridgers.jpp.2017,niel.pre.2018}.
Stochastic effects are, however, lost.

The numerical methods are similar to those employed in the classical case.
If the external field is a plane wave, the lightfront momentum, which satisfies
$\omega_0 \partial_\phi p^- = -(2\alpha/3) |\vec{E}_\perp(\phi)/E_\text{cr}|^2 {p^-}^2 g(\chi)$,
is advanced using Heun's method. The remaining components follow from
the mass-shell condition and
$\omega_0 \partial_\phi (\vec{p}_\perp / p^-) = -e \vec{E}_\perp(\phi) / p^-$,
where we have neglected an additional derivative term that is smaller
by a factor of $\alpha \omega_0/m \ll 1$.
If the field is a focussed Gaussian beam, we employ the Vay push as before,
modified to include RR using the method given in \cite{tamburini.njp.2010}:
the momentum after half a timestep is used to calculate the magnitude of
the RR force $g(\chi) \vec{f}_\text{LL}$, and the associated momentum change
$g(\chi) \vec{f}_\text{LL} \Delta t$ is added to the momentum change induced by
the Lorentz force.

Photons are obtained by pseudorandomly sampling the differential
spectrum \cref{eq:EmissionRate}, as in the quantum case.
However, we do not recoil the electron on emission, as the energy loss
is already accounted for in the equation of motion.


\bibliography{references}

\end{document}